\documentstyle[prl,aps,multicol,psfig]{revtex}
\begin{document}
\draft
\title{Indication of the ferromagnetic instability in a dilute
two-dimensional electron system}
\author{A.~A. Shashkin$^*$ and S.~V. Kravchenko}
\address{Physics Department, Northeastern University, Boston,
Massachusetts 02115}
\author{V.~T. Dolgopolov}
\address{Institute of Solid State Physics, Chernogolovka, Moscow
District 142432, Russia}
\author{T.~M. Klapwijk}
\address{Department of Applied Physics, Delft University of
Technology, 2628 CJ Delft, The Netherlands}
\maketitle
\begin{abstract}
The magnetic field $B_c$, in which the electrons become fully
spin-polarized, is found to be proportional to the deviation of the
electron density from the zero-field metal-insulator transition in a
two-dimensional electron system in silicon. The tendency of $B_c$ to
vanish at a {\it finite} electron density suggests a ferromagnetic
instability in this strongly correlated electron system.
\end{abstract}
\begin{multicols}{2}

At sufficiently low electron densities, an ideal two-dimensional (2D)
electron system becomes strongly correlated, because the kinetic
energy is overpowered by energy of electron-electron interactions
(exchange and correlation energy). The interaction strength is
normally described by the Wigner-Seitz radius, $r_s$, which is equal
to the ratio of the Coulomb and the Fermi energies as calculated for
2D band electrons. In the low-density limit, at $r_s\gtrsim35$, the
ground state of the system is expected to be a Wigner crystal
characterized by spatial and spin ordering \cite{wigner}. The spin
ordering may survive at higher electron densities (lower $r_s$) up to
the threshold determined by ferromagnetic (Stoner) instability
\cite{stoner}. In this strongly-interacting limit ($r_s>>1$), all
results given by theoretical approaches are very approximate, not to
mention that in real 2D electron systems, the influence of disorder
needs to be taken into account, which complicates the problem
drastically. In strongly disordered 2D systems, the influence of the
disorder can dominate the interaction effects leading to a
disorder-driven localization, whereas in the least disordered 2D
systems, the metal-insulator transition may be caused by interaction
effects \cite{rmp}.

In magnetic fields parallel to the 2D electron plane, the spin
effects should dominate the properties of a 2D electron system once
the orbital quantization is quenched. Indeed, the resistance of a 2D
electron gas in silicon metal-oxide-semiconductor field-effect
transistors (MOSFETs) was found to be isotropic with respect to
in-plane magnetic field, $B$, and rise steeply with the field
saturating to a constant value above a critical magnetic field $B_c$
which depends on electron density \cite{sim}. Moreover, an analysis
of Shubnikov-de Haas oscillations in tilted magnetic fields has
established recently that the field $B_c$ corresponds to the onset of
full spin polarization of the electron system \cite{Ok,vit}.

In this paper, we study low-temperature parallel-field
magnetotransport in a 2D electron system in silicon in a wide range
of electron densities, $n_s$. We find that the saturation (or
polarization) magnetic field, $B_c$, is a strictly linear function of
$n_s$: $B_c\propto(n_s-n_c)$ where $n_c$ is the critical electron
density for the $B=0$ metal-insulator transition. Since above $n_c$,
the 2D bandtail of localized electrons is negligibly small, as
inferred from the low-field/low-temperature Hall effect measurements,
vanishing $B_c$ at a finite electron density gives evidence in favor
of the existence of a ferromagnetic transition (at $n_s=n_c$) in this
strongly correlated 2D electron system.

Measurements were made in a rotator-equipped Oxford dilution
refrigerator with a base temperature of $\approx30$~mK on
high-mobility (100)-silicon samples similar to those previously used
in Ref.~\cite{krav00}. The resistance was measured by a standard
4-terminal low-frequency technique. Excitation current was kept low
enough to ensure that measurements were taken in the linear regime of
response. Contact resistances in our samples were minimized by using
a split-gate technique that allows one to maintain a high electron
density in the vicinity of the contacts regardless of its value in
the main part of the sample. In this paper we show results obtained
on a sample with a peak mobility close to 3~m$^2$/Vs at 0.1~K.

Typical curves of the low-temperature magnetoresistance $\rho(B)$ in
a parallel magnetic field are displayed in Fig.~\ref{raw}. The
resistivity increases approximately quadratically with field until it
saturates at a constant value above a certain density-dependent
magnetic field. In the vicinity of the metal-insulator transition,
the magnetoresistance strongly depends on temperature $T$, as was
reported, {\it e.g.}, in Ref.~\cite{sim98}. As one moves away from
the transition, however, this temperature dependence weakens and
eventually disappears at very low temperatures where the metallic
resistivity at $B=0$ saturates and becomes independent of
temperature. Therefore, as $n_s$ is lowered, the low-temperature
limit is realized for progressively narrower initial intervals on the
curve $\rho(B)$. The data discussed in this paper are obtained in
this 
\vbox{
\vspace{-19mm}
\hbox{
\hspace{2mm}
\psfig{file=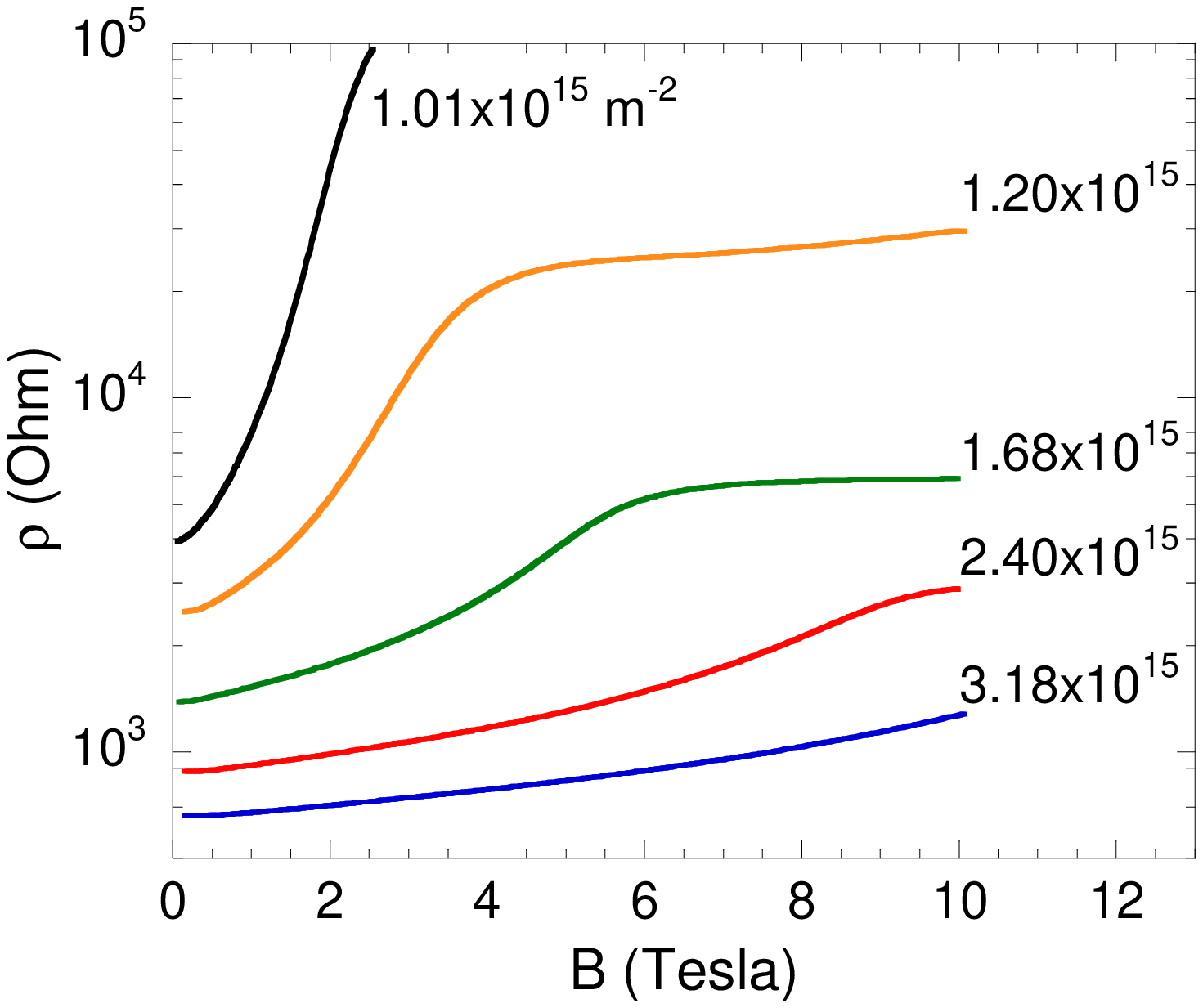,width=2.6in,bbllx=.5in,bblly=1.25in,bburx=7.25in,bbury=9.5in,angle=0}
}
\vspace{-0.3in}
\hbox{
\hspace{-0.15in}
\refstepcounter{figure}
\parbox[b]{3.4in}{\baselineskip=12pt \egtrm FIG.~\thefigure.
Low-temperature magnetoresistance in parallel magnetic
fields at different electron densities above the critical density for
the $B=0$ metal-insulator transition. The lowest density curve is
outside the scaling region as described in the text.\vspace{0.20in}
}
\label{raw}
}
}
low-temperature limit where the magnetoresistance becomes
temperature-independent.

In Fig.~\ref{scal}, we show how the normalized magnetoresistance,
measured at different electron densities, collapses onto a single
curve when plotted as a function of $B/B_c$. The scaling parameter,
$B_c$, has been normalized to correspond to the magnetic field at
which the magnetoresistance saturates (within the accuracy with which
the latter can be determined) and hence full spin polarization of the
electrons is reached \cite{Ok,vit}. The observed scaling is
remarkably good for $B/B_c\le 0.7$ in the electron density range
$1.08\times 10^{15}- 10^{16}$~m$^{-2}$, although with increasing
$n_s$ the scaled experimental data occupy progressively shorter
intervals on the resulting curve. Both at $B/B_c>0.7$ and outside the
indicated range of electron densities, the scaled data start to
noticeably deviate from the universal curve. In particular, the
scaling breaks down when one approaches ($n_s<1.3\,n_c$) the
metal-insulator transition which in this sample occurs at zero
magnetic field at $n_c=8\times 10^{14}$~m$^{-2}$ \cite{rem}. This is
not surprising as the magnetoresistance near $n_c$ depends strongly
on temperature, see above. We note that the observed scaling
dependence is described reasonably well by the theoretical dependence
of $\rho/\rho(0)$ on the degree of spin polarization
$\xi=gm\mu_BB/\pi\hbar^2n_s=B/B_c$ (where both the $g$ factor and the
effective mass $m$ may be enhanced due to interactions, see below) as
predicted by the recent theory \cite{gold} based on the
spin-polarization-dependent screening of a random potential deeply in
the metallic regime, see Fig.~\ref{scal}.

In Fig.~\ref{Bc}, we show the first important and unexpected result
of this paper: with high precision, $B_c$ is proportional to the
deviation of the electron density from its critical value, {\it
i.e.}, to $(n_s-n_c)$, over a wide range of electron densities. In
other words, the field, at which the magnetoresistance saturates,
tends to vanish at $n_c$. We emphasize that our procedure provides
high accuracy for determining the behavior of the field of saturation
with 
\vbox{
\vspace{-19mm}
\hbox{
\hspace{2mm}
\psfig{file=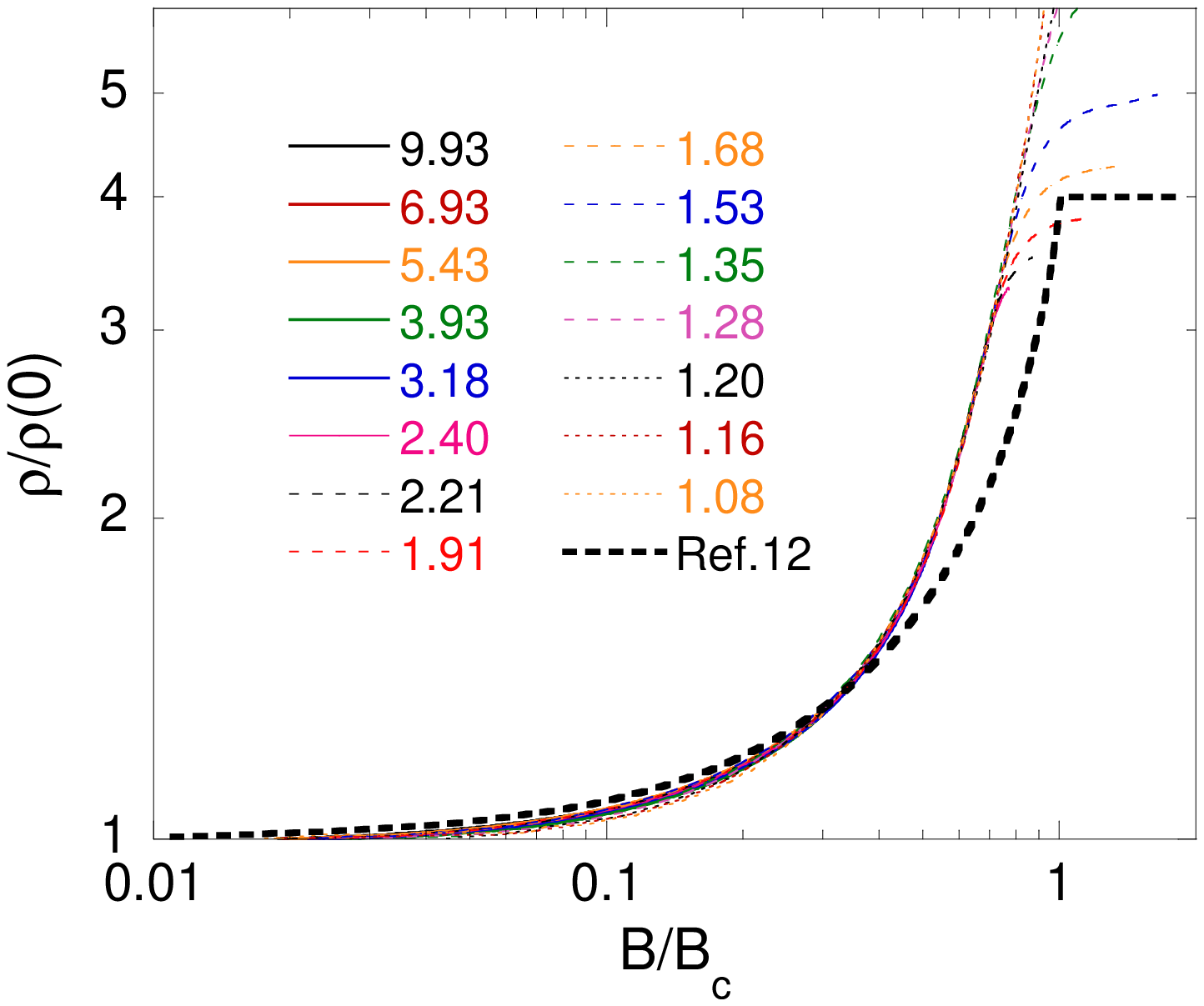,width=2.6in,bbllx=.5in,bblly=1.25in,bburx=7.25in,bbury=9.5in,angle=0}
}
\vspace{-0.3in}
\hbox{
\hspace{-0.15in}
\refstepcounter{figure}
\parbox[b]{3.4in}{\baselineskip=12pt \egtrm FIG.~\thefigure.
Scaled curves of the normalized magnetoresistance at
different $n_s$ vs $B/B_c$. The electron densities are indicated in
units of $10^{15}$~m$^{-2}$. Also shown by a dashed line is the
normalized magnetoresistance calculated in
Ref.~\protect\cite{gold}.\vspace{0.20in}
}
\label{scal}
}
}
electron density, {\it i.e.}, the functional form of $B_c(n_s)$,
even though the absolute value of $B_c$ is determined not so
accurately. Note that at $n_s$ above $2.4\times 10^{15}$~m$^{-2}$,
the saturation of the resistance is not reached in our magnetic field
range; still, the high precision of the collapse of the high-density
experimental curves onto the same scaling curve as the low-density
data allows us to draw conclusion about the validity of the obtained
law $B_c(n_s)$ over a much wider range of electron densities.

The fact that the saturation (or polarization) field tends to vanish
at $n_c$ would be trivial if the density of the delocalized electrons
approached zero at the metal-insulator transition. This crucial point
can be verified with the help of the Hall effect measurements in the
limit of weak magnetic fields and low temperatures \cite{density}. As
seen in the inset to Fig.~\ref{g}, above $n_c$, the electron density
obtained from the Hall effect measurements agrees well with the one
determined from Shubnikov-de Haas oscillations, which shows that in
the metallic phase, the 2D bandtail of localized electron states is
small, {\it i.e.}, all the electrons are delocalized \cite{cap}.
Then, the observed tendency for $B_c$ to vanish at a finite electron
density gives a sign of the long-awaited ferromagnetic instability in
strongly correlated 2D electron systems \cite{afs}.

Subject to the occurrence of a ferromagnetic (Stoner) instability
\cite{stoner} at $n_s=n_c$, it is easy to calculate the behavior of
$B_c$ vs $n_s$ based on the functional form of the exchange and
correlation energy which is valid in the strong-interaction limit.
Apparently, the Fermi energies of a spin polarized and unpolarized 2D
electron system with both interaction and Zeeman energy taken into
account must be equal. Writing the exchange and correlation energy in
the explicit form, we get for the energy difference at the onset of
full spin polarization and at $B=0$

\begin{equation}
\frac{\pi\hbar^2n_s}{2m_b}-\gamma\frac{e^2}{\kappa}n_s^{1/2}-
\frac{1}{2}g_0\mu_BB_c=0,
\end{equation}
\vbox{
\vspace{-19mm}
\hbox{
\hspace{2mm}
\psfig{file=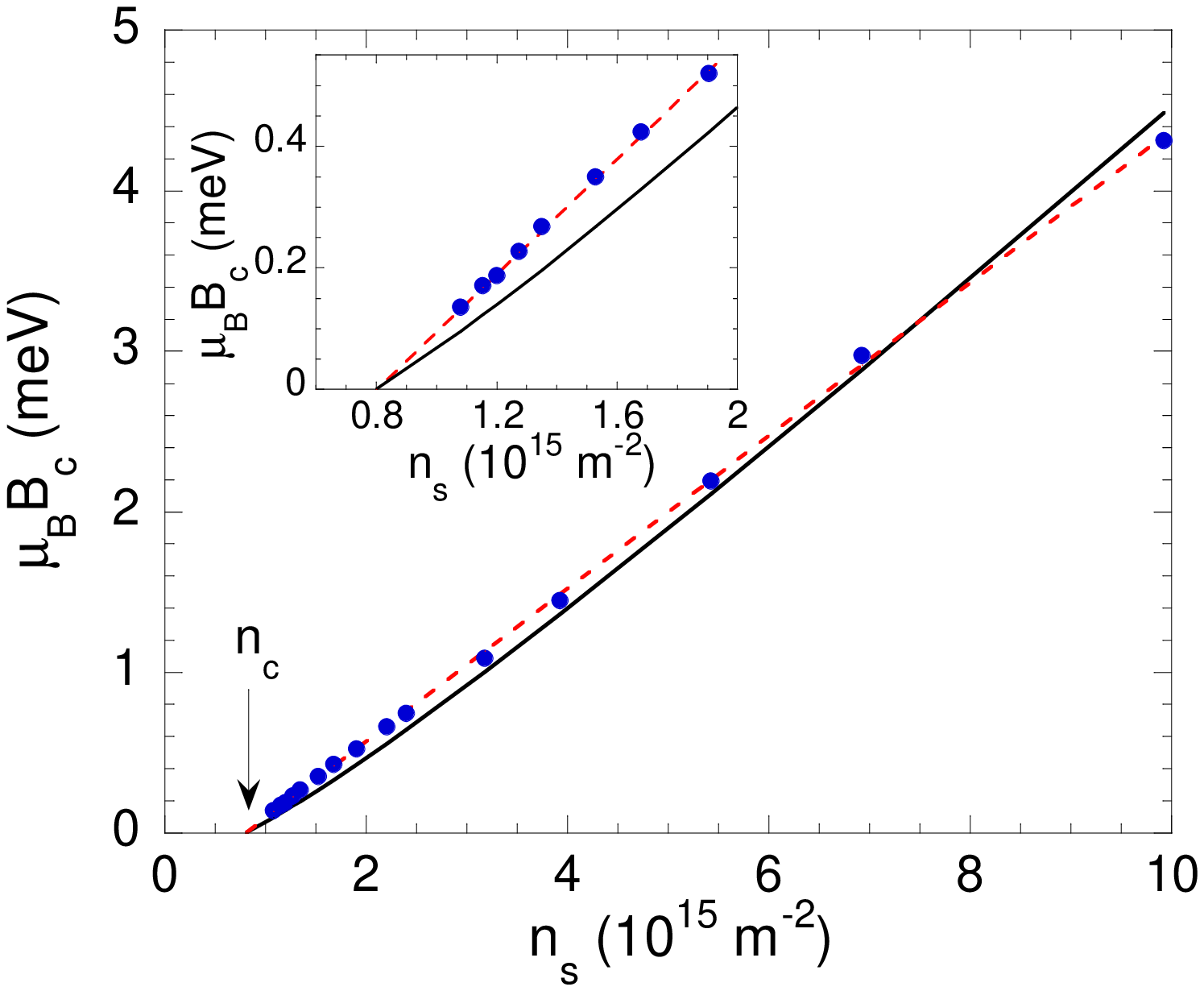,width=2.6in,bbllx=.5in,bblly=1.25in,bburx=7.25in,bbury=9.5in,angle=0}
}
\vspace{-0.3in}
\hbox{
\hspace{-0.15in}
\refstepcounter{figure}
\parbox[b]{3.4in}{\baselineskip=12pt \egtrm FIG.~\thefigure.
Dependence of the field $B_c$ on electron density. The
dashed line is a linear fit which extrapolates to the critical
electron density for the $B=0$ metal-insulator transition. The fit
using Eq.~(1) with $\gamma=0.095$ is shown by a solid line. A
close-up view of the region near $n_c$ is displayed in the
inset.\vspace{0.20in}
}
\label{Bc}
}
}
where $\hbar$ is the Planck constant, the band mass $m_b=0.19\,m_0$
($m_0$ is the free electron mass), $e$ is the electron charge,
$\kappa$ is the dielectric constant, the Land\'e $g$ factor in bulk
silicon is equal to $g_0=2$, and $\mu_B$ is the Bohr magneton. The
first term in Eq.~(1) is given by the difference of the bare Fermi
energies which differ by a factor of two originating from the spin
degeneracy (valley degeneracy being taken into account). The
interaction (second) term contains the unknown numerical factor
$\gamma$ which is positive since the exchange and correlation energy
is always negative and its absolute value enhances with spin
polarization. Note that the interactions are expected to lead to
corrections to both $m_b$ and $g_0$; the problem of the
interaction-enhanced effective mass $m$ and $g$ factor lacks a
definite answer so far. Note also that the drastic enhancement of the
product $gm$ (Fig.~\ref{g}) determined from the data in
Fig.~\ref{Bc}, based on the above form of the spin polarization
parameter $\xi$ is consistent with the $gm$ value obtained from the
analysis of Shubnikov-de Haas oscillations in dilute silicon MOSFETs
\cite{Ok,fang,krav}.

As seen from Fig.~\ref{Bc}, the resulting dependence $B_c(n_s)$ with
the factor $\gamma=0.095$ determined from the condition $B_c(n_c)=0$
describes the experimental finding well enough. Formally, the best
description of the experimental data might be obtained with a bit
smaller $\gamma=0.090$, which corresponds to the expected
ferromagnetic transition point just below $n_c$, at $n_s=7\times
10^{14}$~m$^{-2}$. Although the accuracy of the $B_c$ normalization
hardly allows one to do so, the conclusions of the paper would still
hold for that case. We emphasize, however, that the $B_c(n_s)$
dependence given by Eq.~(1) is weakly superlinear, whereas our
experiments yield strictly linear dependence. Clearly, theoretical
efforts are needed to explore a possible relationship between the
Stoner instability and the metal-insulator transition in 2D.

In summary, we have studied the low-temperature magnetoresistance in
parallel magnetic fields 
%
in a wide 
\vbox{
\vspace{-19mm}
\hbox{
\hspace{2mm}
\psfig{file=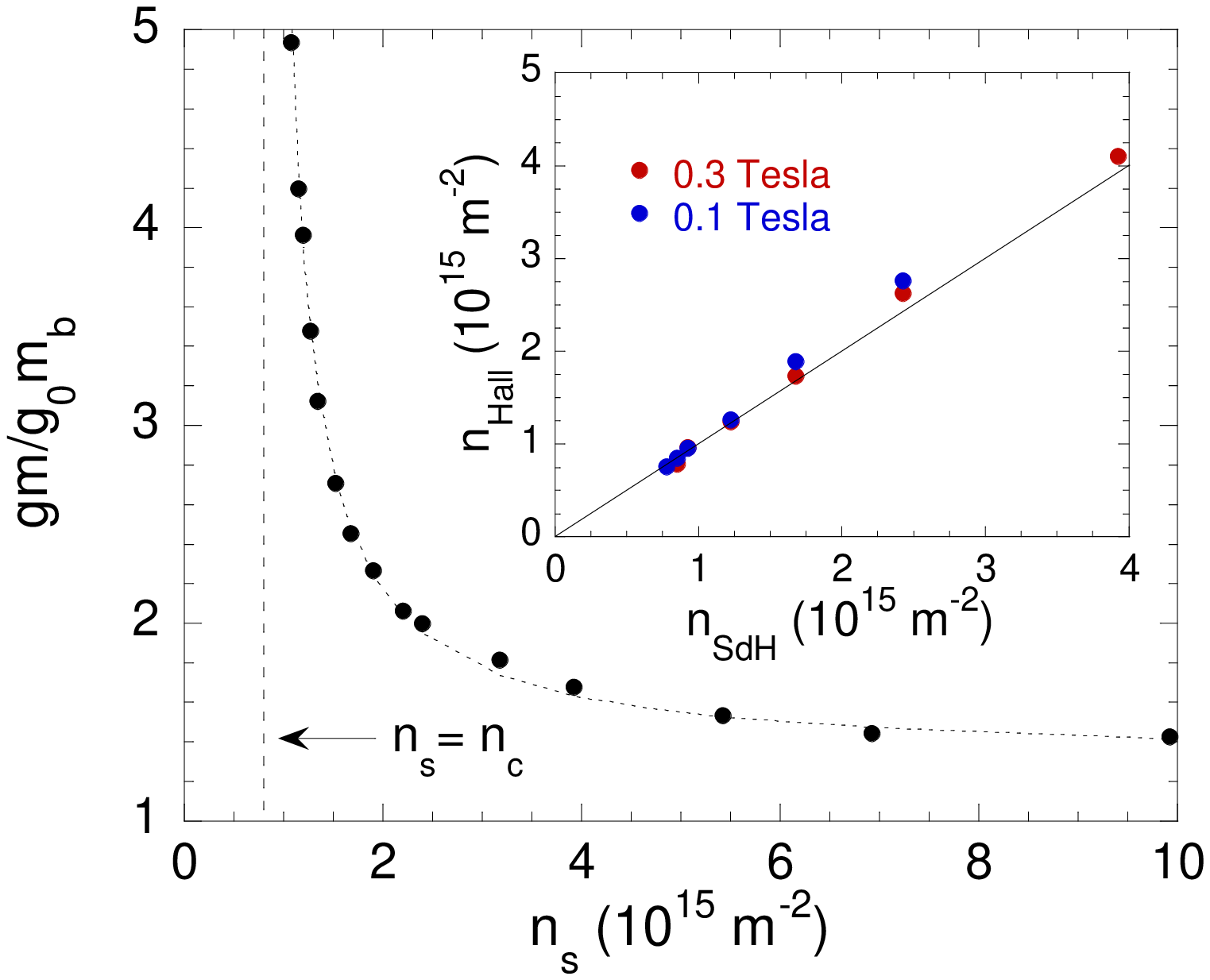,width=2.6in,bbllx=.5in,bblly=1.25in,bburx=7.25in,bbury=9.5in,angle=0}
}
\vspace{-0.3in}
\hbox{
\hspace{-0.15in}
\refstepcounter{figure}
\parbox[b]{3.4in}{\baselineskip=12pt \egtrm FIG.~\thefigure.
The product $gm/g_0m_b$ vs $n_s$ obtained from the data in
Fig.~\ref{Bc}, based on the above form of the spin polarization
parameter $\xi$. The dashed line is a guide to the eye. The inset
compares the electron densities determined from Shubnikov-de Haas
oscillation and weak-field low-temperature Hall measurements. For the
latter, averages were taken for opposite magnetic field
directions.\vspace{0.20in}
}
\label{g}
}
}
range of electron densities above the critical electron 
density for the zero-field
metal-insulator transition, $n_c$. 
The normalized magnetoresistance is found to scale with $n_s$
defining the scaling parameter, $B_c$, 
that corresponds to the magnetic field in which the full spin 
polarization is achieved.  Over a wide range of electron densities, 
this scaling parameter changes in precise accordance with the relation 
$B_c\propto(n_s-n_c)$.
Although in the metallic regime, the bandtail of the
localized electrons has been found negligibly small, $B_c$ tends to
vanish at a finite electron density. This gives evidence in favor of
the existence of a ferromagnetic transition in this strongly
correlated 2D electron system.

Recently, a conclusion about possible ferromagnetism in a dilute 2D
electron system has been made by another experimental group 
\cite{myriam}.

We gratefully acknowledge discussions with I.~L. Aleiner, A.~H.
Castro Neto, C. Chamon, B.~I. Halperin, D. Heiman, A.~I. Larkin,
M.~P. Sarachik, and S.~A. Vitkalov. This work was supported by NSF
grants DMR-9803440 and DMR-9988283, RFBR grant 01-02-16424, and Sloan
Foundation.





\end{multicols}

\begin{references}
\bibitem{rem} The $B=0$ metal-insulator transition point has been
determined using the criterion of a sign change of the derivative
${\rm d}\rho/{\rm d}T(n_s)$ as well as the alternative methods of
vanishing activation energy and vanishing nonlinearity of
current-voltage curves as extrapolated from the insulating phase
\cite{sh94}. For our case all of the definitions are equivalent
\cite{phb}.
\bibitem{density} The delocalized electron density could in principle
be affected by the application of an insufficiently weak magnetic
field \cite{tail}. Besides, at relatively high temperatures
\cite{sar00}, it may depend on temperature \cite{pud}.
\bibitem{cap} The electron density variation due to inhomogeneities
in the sample can be estimated from the width of the
magnetocapacitance minima in a quantizing magnetic field; it is about
$4\times 10^{13}$~m$^{-2}$ at $n_s\sim 2\times 10^{15}$~m$^{-2}$ in
our samples.
\end{references}
\end{document}